\documentclass[preprint,12pt]{elsarticle}

\usepackage{epsfig}
\usepackage{graphicx}
\usepackage{amsmath}
\usepackage{multirow}
\usepackage{setspace}
\usepackage{appendix}

\begin{document}

\begin{frontmatter}

\title{ Aggregate download throughput for TCP-controlled long file transfers in 
a WLAN with multiple STA-AP association rates }
\date{}
\author{Pradeepa BK and Joy Kuri \\ Centre for Electronics Design and Technology, \\
Indian Institute of Science, Bangalore. India. \\
bpradeep@cedt.iisc.ernet.in, kuri@cedt.iisc.ernet.in
}
\begin{abstract}
We consider several WLAN stations associated at rates $r_1$, $r_2$, ..., $r_k$ 
with an Access Point. Each station (STA) is downloading a long file from a local 
server, located on the LAN to which the Access Point (AP) is attached,
using TCP. We assume
that a TCP ACK will be produced after the reception of $d$ packets at an STA. We 
model these simultaneous TCP-controlled transfers using a semi-Markov process. Our 
analytical approach leads to a procedure to compute aggregate download, 
as well as per-STA throughputs,
 numerically, and the results match simulations very well.
\end{abstract}
\end{frontmatter}

\section{Introduction}\label{sec:Introduction}
IEEE 802.11a/b/g/n based Wireless Local Area Networks
(LANs) in ``infrastructure mode'' are very common
 in many places. In this paper, we are concerned with an analytical model for
 evaluating  the performance of TCP-controlled downloads in a WLAN. 
``TCP'' is the Transmission Control Protocol, which is regarded as the
workhorse of the Internet; numerous applications, including web browsing, file 
transfer, and secure e-commerce, rely on TCP as the transport protocol.
The system we consider is shown in Figure~\ref{fig:AP_STA}.
A detailed 
analysis of the aggregate throughput of TCP-controlled file downloads for a ``single 
rate'' Access Point (AP) is given in Kuriakose et al. \cite{astn_model:Kuriakose}; in this, all STAs
are assumed to be associated with the AP at the same rate.
In practice, many
data rates are possible and hence considering multiple rates is important. The 
aggregate download throughput is evaluated for the \emph{two rates} case in 
Krusheel and Kuri \cite{astn_model:Krusheel}. In this paper, we consider an arbitrary but
finite number $k$ of possible rates of 
association between stations (STA) and a single AP.

We are motivated to study an analytical model because of the improved 
understanding that it leads to, and the useful insights that it can provide. 
Closed-form expressions or numerical calculation procedures are helpful because
other features and capabilities can be built upon them. One possible application, 
which we are studying now, is to utilize the results reported here in devising a 
better AP-STA association policy.

Our approach is to model the number of STAs with TCP Acknowledgements (ACKs)
in their Medium Access Control (MAC) queues as an 
embedded discrete time Markov chain (DTMC), embedded at the instants of successful
transmission events. We consider a successful transmission from the AP as a reward.
 This leads to viewing the aggregate TCP throughput in the framework of Renewal 
Reward theory given in Kumar \cite{astn_model:ak_notes}.

Almost the entire existing literature considers a \emph{single} rate of association only.
This is rather limiting, because in practice, it is extremely likely
that a WLAN will have STAs associated at a number of rates allowed by the
technology (for example, one of 6 Mbps, 12 Mbps, 18 Mbps, 24 Mbps, 30 Mbps,
36 Mbps, 48 Mbps or 54 Mbps in 802.11g, and one of 1 Mbps, 2 Mbps, 
5.5 Mbps or 11 Mbps in 802.11b). A first step towards considering multiple
association rates was taken in Krusheel and Kuri \cite{astn_model:Krusheel}, but there, only
2 possible association rates were considered. In this paper, \emph{any} number
of association rates is allowed. Because of this, our model is applicable to
any variant of WLAN technology, for example: 802.11a/b/g/n.

The contributions of this paper are as follows. We present a
model for analyzing the performance of TCP-controlled file transfers 
with $0 < k < \infty$ rates of association. This generalizes earlier work.
Secondly, our model incorporates TCP-specific aspects like ``delayed ACKs;''
this is a technique to reduce the frequency of TCP ACK generation by a 
TCP receiver. In most implementations, a TCP receiver generates a TCP ACK
for every $2^{\mathrm{nd}}$ TCP packet received; our model is general, and
considers that one TCP ACK is generated for every $d$ TCP packets.
Our analytical results are in excellent agreement with simulations, with the
discrepancy being less than $1$\% in all cases.

The paper is organized as follows: In Section~\ref{sec:Related_Work}, related
works are discussed. In Section~\ref{sec:System_Model}, we state the
assumptions in first part and then present our analysis. In Section~\ref{sec:Evaluation},
we present performance evaluation results. In Section~\ref{sec:Discussion} we discuss
the results. Finally, the paper is concluded in Section~\ref{sec:Conclusion}.

\section{Related Work}\label{sec:Related_Work}

The literature on throughput modelling in a WLAN can be classified into 
several groups depending on the approach.

In the first group, all WLAN entities (STAs and the AP) are assumed to
be \emph{saturated}, \textit{i.e.}, each entity is backlogged permanently.
Bianchi \cite{astn_model:bianchi}, Kumar et al. \cite{astn_model:kumar} and
Cali et al. \cite{astn_model:Cali} consider this saturated traffic model. However, 
our interest is in modelling aggregate \emph{TCP} throughput, and the
saturated traffic model does not capture the situation well.

\begin{figure}[!hH]
\centering
\includegraphics[scale=0.45]{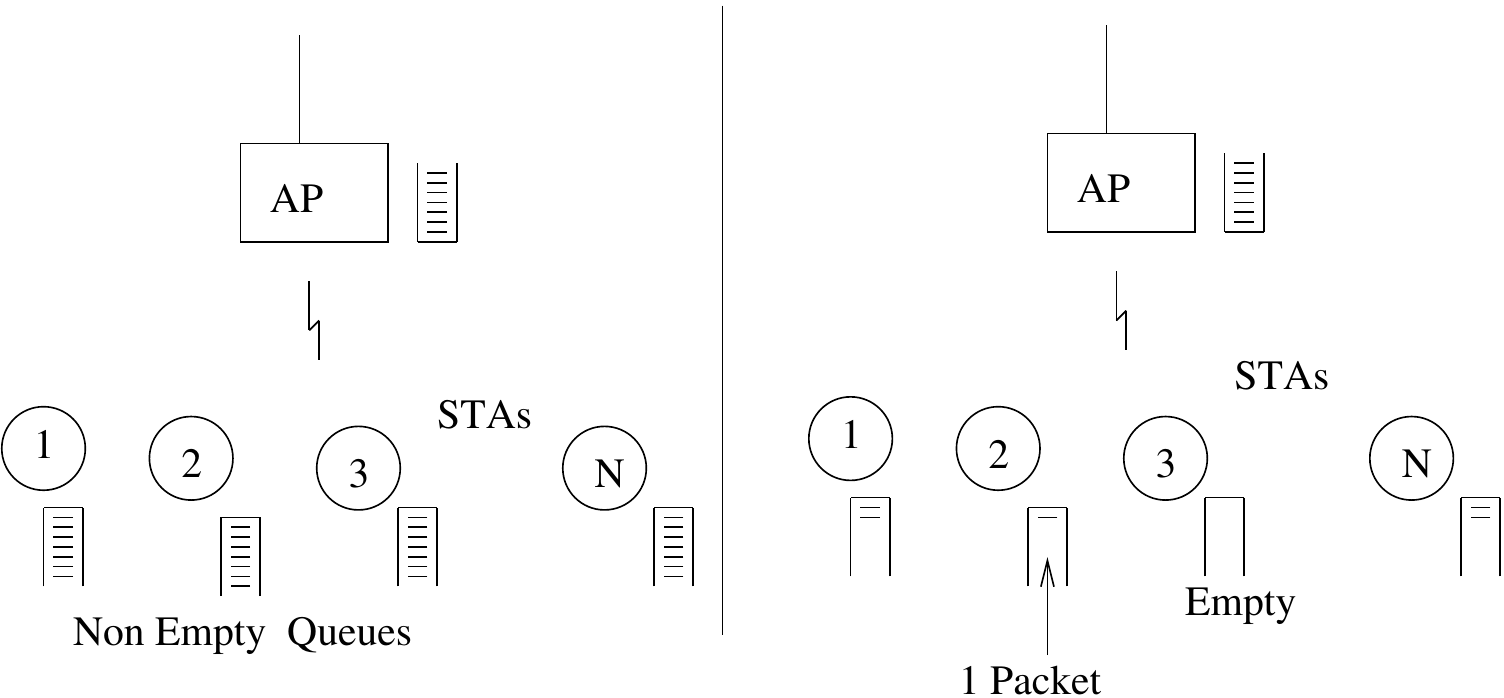} 
\caption{Two scenarios are shown: the one on the left with saturated traffic and the
one on the right with unsaturated traffic. The queues of the AP and STAs are
shown next to them. In the saturated traffic scenario, both the AP and the STAs
have packets in their queues all the time. In the unsaturated traffic scenario which
we are addressing, the AP has packets to transmit all the time but the
MAC queues at STAs are empty most of the time.}
\label{fig:unsaturated}
\end{figure}

To see why unsaturated traffic makes a difference, we consider Figure 
\ref{fig:unsaturated}. The left part shows a saturated traffic scenario,
where all WLAN entities have packets to transmit; therefore, $(N+1)$ entities
contend for the channel. The right part shows the situation with TCP in the
picture. Essentially, for many TCP connections, the entire window of packets
sits at the AP, leaving the corresponding STAs with nothing to send. This means
that the number of contending WLAN entities is much smaller as mentioned in
Kuriakose et al. \cite{astn_model:Kuriakose}. This indicates why approaches relying on a model
with saturated nodes are inadequate.

The second group considers TCP traffic. Kuriakose et al. \cite{astn_model:Kuriakose}
propose a model for TCP-controlled file downloads
in a \emph{single rate} WLAN; \textit{i.e.,} one in which all STAs are
associated with a single AP at the \emph{same} rate.
Bruno et al. \cite{astn_model:Bruno1}, \cite{astn_model:Bruno2} \cite{astn_model:Bruno3},
\cite{astn_model:Bruno4},  \cite{astn_model:Bruno5}, and Vendictis et al. \cite{astn_model:Vendictis}
generalize this and analyze TCP-controlled file uploads as well as
downloads; however, it is assumed again that all STAs are associated at
the same rate.
Similarly, Yu et al. \cite{astn_model:Yu} 
provide an analysis for a given number of STAs and a maximum TCP receive window 
size by using the well-known $p$-persistent model proposed in 
Cali et al. \cite{astn_model:Cali}. 
As noted above, these papers analyze TCP-controlled file transfers (in
some cases UDP traffic is allowed as well) but limit themselves  
to a single rate of association.



In the third category, Bharadwaj et al. \cite{astn_model:Onkar} consider
\emph{finite} AP buffers, in contrast to
the previous two, where AP buffers were assumed to be infinite;
however, the single rate assumption is retained.

The three groups mentioned above focus on \emph{long} file transfers, where the
TCP sender is assumed to have a file that is infinite in size.  
Miorandi et al. \cite{astn_model:Miorandi}
model a different situation motivated by web browsing over a WLAN.
A queuing model is proposed to compute the mean session delay for short-lived TCP
flows. The impact of maximum TCP congestion window size on this delay
is studied as well.

Even though a fair amount of work modelling TCP-controlled transfers has been
done, we are unaware of any work that allows \emph{multiple} AP-STA association
rates. Clearly, this is the situation observed most often in practice, where the
distance between the AP and a STA governs the rate of association. In this paper,
we consider an arbitrary (but finite) number of rates of association between
STAs and the AP; to the best of our knowledge, this is the first paper to consider
this general model.
\section{System Model}\label{sec:System_Model}
\subsection{Assumptions}\label{sec:Assumptions}
\begin{figure}[!hH]
\centering
\includegraphics[scale=0.5]{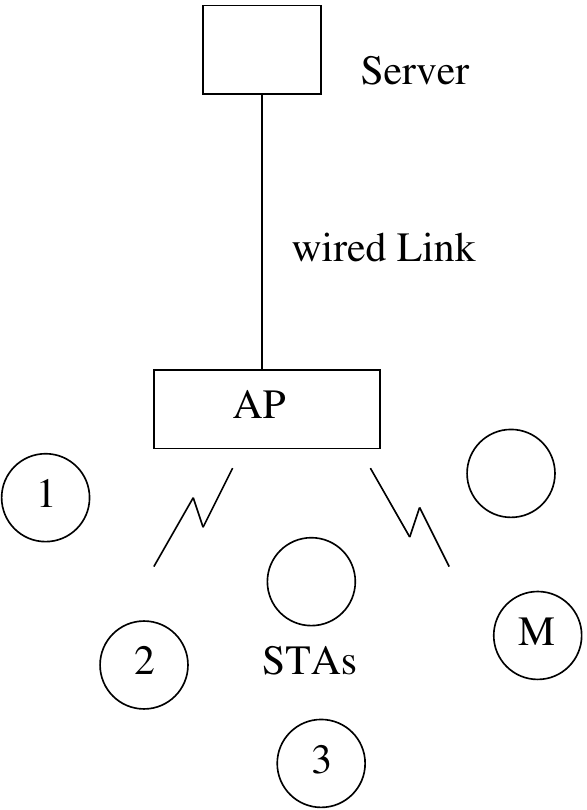} 
\caption{STAs downloading files from a server through an AP.}
\label{fig:AP_STA}
\end{figure}

We consider $M$ stations associated with an AP as shown in Figure 
\ref{fig:AP_STA}. All the nodes contend for the channel using the DCF mechanism
as given in IEEE 802.11a/b/g/n. The stations are associated with the AP at $k$ 
different physical rates ($m_1$ STAs at rate $r_1$, $m_2$ STAs at rate $r_2$, $\ldots$ $m_k$ 
STAs at rate $r_k$). We assume that there are no link errors. This is not merely
a simplifying assumption; the ``auto rate fallback'' mechanism, implemented widely
in STAs and APs, is intended to ensure that we have an \emph{error-free but lower rate}
channel rather than a higher rate but error-prone channel. Thus, our assumption
of no link errors is consistent with the usual mode of WLAN operation.

Packets in the medium are lost only due to collisions. Each station has a single TCP connection 
to download long files from the server and all TCP connections have equal window 
sizes\footnote {This can be generalized, as in Pradeepa and Kuri \cite{astn_model:pradeep_kuri2}.}. 
The AP transmits TCP packets for the stations and the stations return 
TCP-ACK packets. Further, we assume that the AP uses the RTS-CTS mechanism while 
sending packets to stations and stations use basic access to send ACK packets
(RTS: Request to Send, CTS: Clear to Send; these are control packets that
reserve the wireless medium for the subsequent long data packet).
In IEEE 802.11 WLANs, the \textrm{RTS Threshold} parameter determines whether
the RTS-CTS exchange will precede a packet transmission. In most operational WLANs,
the RTS Threshold is set such that TCP data packets are larger than the RTS Threshold,
and hence sent after RTS-CTS exchange, while TCP ACK packets are smaller than the
RTS Threshold, and hence sent without RTS-CTS exchange (the latter is referred to
as ``basic access'').

Upon reception of $d$ data packets, a STA generates an ACK packet and
it is enqueued at the MAC layer for transmission. We assume that all nodes have
sufficiently large buffers, so that packets are not lost due to buffer overflow. 
Also, TCP timeouts do not occur. TCP start-up transients are ignored by considering 
all connections to be in Congestion Avoidance. For long file transfers (which we
are considering in this paper), this is a reasonable assumption because the 
initial start-up phase of a TCP connection lasts for a time that is completely
negligible compared to the connection lifetime; the TCP connection moves quickly
to the Congestion Avoidance phase and remains there.
The value of RTT is very small, since
files are downloaded from a server located on the LAN as shown in Figure~\ref{fig:AP_STA}.

Thus, several TCP connections exist simultaneously and all STAs with TCP ACK
packets, and the AP (which is full of TCP data packets for the STAs),
contend for the channel. Since no 
preference is given to the AP, and it has to serve all STAs, the AP becomes a 
bottleneck, and it is modelled as being backlogged permanently. The aggregate
throughput of the AP is shared equally among all $M$ stations. 

\subsection{Analysis}\label{sec:Analysis}
 Let $m_i $ be the number of stations associated with the AP at 
the physical transmission rate $r_i$, where $i$ $ \in \lbrace 1,2, \ldots k \rbrace $ 
with $r_1  >  r_2  > ... r_k $. Given that the AP wins the channel,
the conditional probability that it sends a TCP data 
packet to a station at rate $ r_i $ is $ p_i $. We assume that $m_i$ is large. 
Our results will show that $m_i \geq 3$ or $4$, $1 \leq i \leq k$  (with $k \geq 4$) 
suffices for the analysis to be applicable.

\begin{figure}[!hH]
\centering
\includegraphics[scale=0.45]{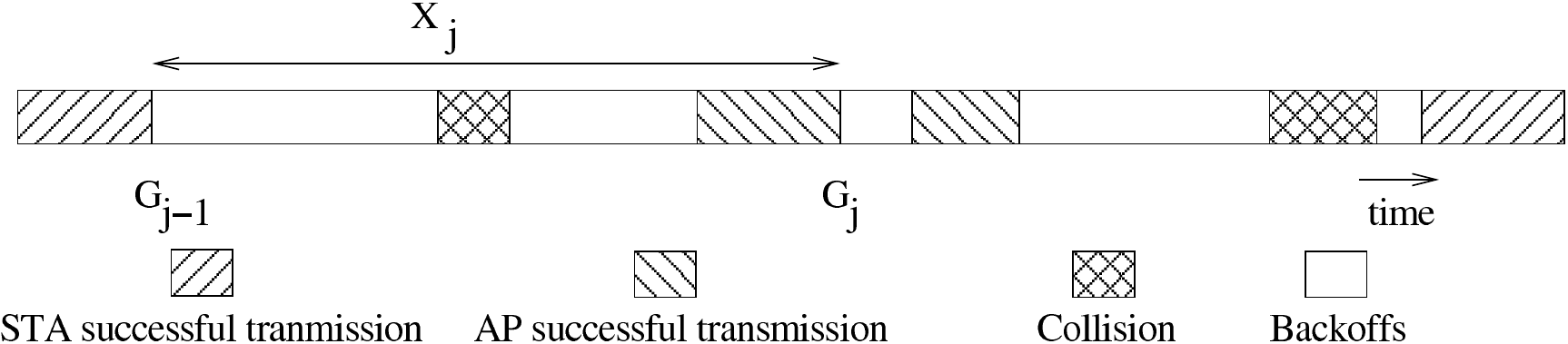} 
\caption{ A possible sample path of events in WLAN shows the backoffs and the 
channel activity. }
\label{fig:channel_activity}
\end{figure}

 Figure~\ref{fig:channel_activity} shows a possible sample path of the events on
 the WLAN channel. The random epochs $ G_j $ indicate the end of the $ j^{th} $ 
successful transmission from either the AP or one of the stations. We observe 
that most STAs have empty MAC queues, because, in order for many STAs to have 
TCP-ACK packets, the AP must have had a long run of successes -- and this is 
unlikely because no special preference is given to the AP. So when the AP succeeds
 in transmitting, the packet is likely to be for a STA with an empty MAC queue. 

At epoch $G_j$,
let $ S_{i,j} $ be the number of stations at rate $ r_i $, ready with an ACK. 
Let $ \sum_{i=1}^{k} S_{i,j} = N_j $ be the number of nonempty STAs. If there
 are $N$ nonempty STAs and a nonempty AP, each nonempty WLAN entity attempts to 
transmit with probability $ \beta_{(N+1)} $ as in Kumar et al. \cite{astn_model:kumar}, where
$\beta_{N+1}$ is the attempt probability with $(N+1)$ \emph{saturated} entities. 
It can be seen that 
$( S_{1,j}, S_{2,j},..., S_{k,j} )$ evolves as a Discrete Time Markov Chain 
(DTMC) over the epochs $ G_j $. This allows us to consider 
$((S_{1,j}, S_{2,j},..., S_{k,j} ) , G_j) $ as a Markov Renewal Sequence, and 
$( S_{1}(t), S_{2}(t),..., S_{k}(t) )$ as a semi-Markov process.

\begin{figure}[!hH]
\centering
\includegraphics[scale=.4]{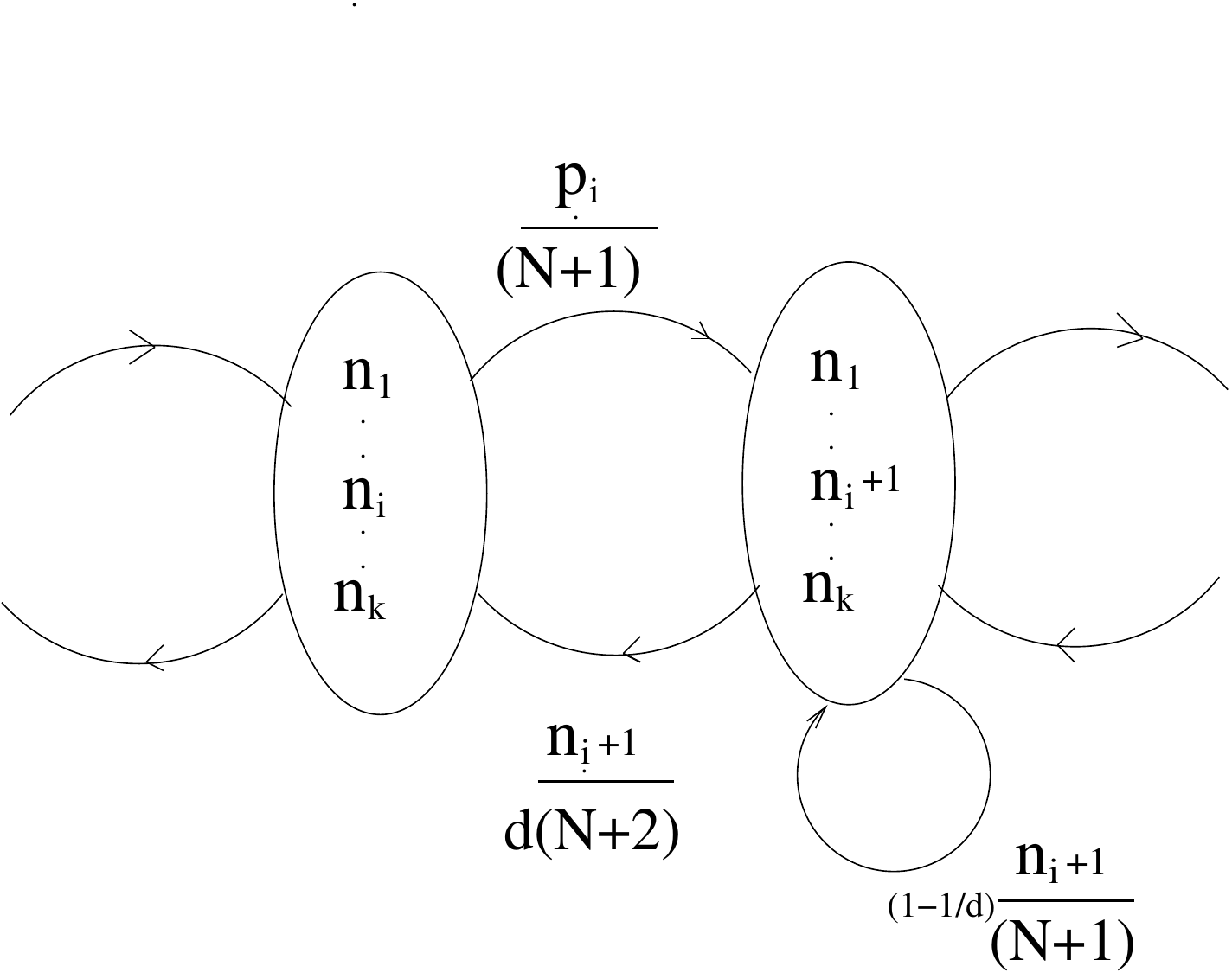}
\caption{Embedded Markov chain formed by the AP and $ n_1 + n_2 + ... + n_k = N $
stations associated with the AP at $ k $ different data rates. }
\label{fig:MarkovChain}
\end{figure}

We have a multidimensional DTMC which is shown in Figure \ref{fig:MarkovChain}; 
transition probabilities are indicated as well. A STA generates a TCP ACK 
after receiving $d$ TCP packets; this is incorporated in our model in the 
following way. When the DTMC state is $(n_1, n_2, \ldots, (n_{i}+1), \ldots, n_k)$,
there are $(n_1 + n_2 + \ldots (n_i + 1) + \ldots + n_k + 1) = N+2$ backlogged
WLAN entities (including the AP); so, the probability that a station at rate
$r_i$ wins the channel is $\frac{n_i + 1}{N+2}$, since each entity is equally
likely to win the contention. Further, given that a STA
at rate $r_i$ wins the channel, the conditional probability that it will
generate a TCP-ACK is $\frac{1}{d}$.

By inspection, we can say that 
the DTMC is irreducible; further, the Detailed Balanced Equation holds for a properly 
chosen set of equilibrium probabilities. The Detailed Balance Equation (DBE) is

\begin{equation}
\begin{split}
\forall n=(n_1, n_2, \ldots, n_k): & n_i \geq 0, 1 \leq i \leq k, \\
 \pi(n_1,n_2,...n_i ,...n_k ) \frac{p_i}{(N+1)} 
= & \frac{1}{d} \pi(n_1,n_2,...(n_i +1),...n_k )\frac{(n_i+1)}{(N+2)} \\
 1 \leq i \leq k
\label{eq:DBE}
\end{split}
\end{equation}

Here $ \pi(n_1,n_2,...n_i ,...n_k ) $, $ n_1, n_2, ... n_k  \in \lbrace 
0,1,2,... \rbrace $ is the stationary distribution of the DTMC. From the set 
of equations given in \eqref{eq:DBE} and 
\begin{equation*}
\sum _{n_1=0} ^{ \infty } \sum _{n_2=0} ^{ \infty }... \sum _{n_k=0} ^{ 
\infty } \pi(n_1,n_2,...n_i ,...n_k )  = 1 ,
\end{equation*}
the stationary distribution is
\begin{equation}
\begin{split}
 \pi(n_1,n_2,...n_i ,...n_k ) = (N+1) \Pi _{i=1} ^{k} \frac{ d^{n_i} (p_i)^{n_i} 
}{(n_i!)} * \frac{1}{(2e)}  
\end{split}
\label{eq:stationary_dist}
\end{equation}	

To obtain the throughput, we use Markov regenerative analysis,
culminating in the Renewal 
Reward Theorem given in Wolff \cite{astn_model:Wolff}, Kumar \cite{astn_model:ak_notes}.
For a given state $(n_1,...n_k)$, successive
entries into state $(n_1,..., n_k)$ form renewal epochs. To obtain the mean
time between successive entries (the mean renewal cycle lengths),
we obtain the mean sojourn time in the state $(n_1,...n_k)$.
	
Let $ X $ be the sojourn time in a state $ ( S_{i,j}, ... S_{k,j} ) $. 
Conditioning on various events (idle slot, collision or successful transmission)
that can happen in the next time slot, the following expression for the mean 
cycle length can be written down: \\
\begin{equation}
\begin{split}
E_{n_1..n_k}X & = P_{idle}(\delta + E_{n_1..n_k}X)  + \Sigma_{i} ( P^{r_i} 
_{sAP}  T^{r_i}_{sAP} )
 + \Sigma_{i} ( P^{r_i} _{c} ( T^{r_i} _{c} + E_{n_1..n_k}X ) )\\
& +\Sigma_{i} ( P^{r_i} _{sSTA}  T^{r_i}_{sSTA} ) 
 + \Sigma_{i} ( P^{r_i} _{cSTA} ( T^{r_i} _{cSTA} + E_{n_1..n_k}X ) ) 
\label{eq:enx11}
\end{split}
\end{equation}
In the above expression \eqref{eq:enx11}, $ P_{idle} $ is the probability of the
slot being idle, $ P^{r_i}_{sAP} $ is the probability that the AP wins the 
contention and transmits the data packet at rate $ r_i$ and $P^{r_i}_{sSTA}$ 
is the probability that a STA associated at rate $r_i$ wins the channel
(``s'' in the suffix stands for ``success''). Correspondingly, the 
conditional expected sojourn times in state $(n_1..n_k)$, given the events 
are, respectively,
$(\delta + E_{n_1..n_k}X)$, $T^{r_i}_{sAP}$ and $T^{r_i}_{sSTA}$.
Detailed expressions for these quantities are provided in the Appendix.

The $3^{\mathrm{rd}}$ and $5^{\mathrm{th}}$ terms on the right side of \eqref{eq:enx11}
correspond to collision events.
The third term in \eqref{eq:enx11} arises when the AP transmits a TCP 
data packet to a station at rate $ r_i $ and some 
other stations are involved in a collision; in other words, the third term captures
the situations in which the AP is involved in a collision.
The fifth term in \eqref{eq:enx11} captures collision events in which the
AP is \emph{not} involved; we have a
STA transmitting a TCP ACK packet to the AP 
at rate $ r_i $ and one or more other STAs transmitting simultaneously. The 
various probabilities have been
obtained by using the attempt probability $\beta _{N+ 1}$, 
when there are $(N + 1)$ contending nodes.
From Equation \eqref{eq:enx11} we have $ E_{n_1..n_k}X = $
\begin{equation}
\frac{ P_{idle} + \sum P^{r_i} _{sAP} T^{r_i} _{sAP} + \sum 
P^{r_i} _{c}T^{r_i} _{c}  +\sum P^{r_i} _{sSTA} T^{r_i} _{sSTA} +\sum P^{r_i} 
_{cSTA} T^{r_i} _{cSTA}   }{1- P_{idle} -\sum P^{r_i} _{c}
-\sum P^{r_i} _{cSTA} }
\label{eq:enx}
\end{equation}

We are 
interested in finding the long run time average of successful transmissions 
from the AP. We obtain this by applying the Renewal Reward 
Theorem of Wolf \cite{astn_model:Wolff}. To get the mean renewal cycle length, we can 
use the mean sojourn time given in Equation~\eqref{eq:enx} and use Theorem~5.3 
in \cite{astn_model:ak_notes}.
The mean reward in a cycle can be obtained
 as follows. A reward of 1 is earned when the AP transmits a TCP data packet 
successfully by winning the channel. The probability of the AP winning the 
channel is  $ \frac{1}{(n_1 + n_2 + ... n_k + 1)}$. 
Similarly, a reward of 0 is earned with
probability $ ( 1- \frac{1}{(n_1 + n_2+ ... n_k + 1)} )$. Therefore, the expected
reward is  $ \frac{1}{(n_1 + n_2 + ... n_k + 1)}$.

Putting all this together, the aggregate TCP throughput can be 
calculated as \cite{astn_model:ak_notes} \\
\begin{equation}	
 \Phi_{AP-TCP} = \frac{\Sigma^{\infty}_{n_1 =0} \Sigma^{\infty}_{n_2 =0} ... 
 \Sigma^{\infty}_{n_k =0} \pi (n_1... n_k)\frac{1}{n_1+...+n_k+1} 
}{\Sigma^{\infty}_{n_1 =0} \Sigma^{\infty}_{n_2 =0} ... \Sigma^{\infty}_{n_k 
=0}  \pi (n_1... n_k)  E_{n_1,..n_k}X}  
\label{eq:ap_thpt}
\end{equation}
\section{Evaluation}\label{sec:Evaluation}
To verify the accuracy of  the model, we performed experiments using the Qualnet~4.5
network simulator \cite{astn_model:Qualnet}. We considered 802.11b physical data 
rates: 1Mbps, 2Mbps, 5.5Mbps and 11Mbps; higher rates correspond to
smaller distances between the STAs and the AP.
In~Table \ref{table:aps_stas}, results are given for a few cases of this 
multirate scenario. For example, the first row considers a total of 10 STAs
associated with the AP, out of which 2 STAs are associated at 11~Mbps and 2~Mbps respectively,
while 3 STAs are associated at 5.5~Mbps and 1~Mbps, respectively.
The values of $p_i$ 
in Equation \eqref{eq:stationary_dist} are calculated by using the number of STAs 
associated with the AP. For example, in the first row in Table 
\ref{table:aps_stas}, $ p_1 $ (for 11 Mbps) is $ \frac{2}{10}$. 
\begin{table}[!hH]
\centering 
\begin{tabular}{| c |c | c| c | c | c | c | c|} 
\hline 
 & \multicolumn{4}{|c|}{No. of STAs at rate (Mbps)} & \multicolumn{3}{|c|}{ Aggregate Throughput (Mbps)} \\ \hline
 M & 11 & 5.5 & 2  & 1 &  Analysis & Simulation & Error \%  \\  [0.5ex]
\hline 
\multirow{2}{*}{10} 	& 2 & 3 & 2 & 3 & 1.0569 & 1.0492 & 0.80 \\ 
			& 1 & 2 & 3 & 4 & 0.8397 & 0.8329 & 0.81 \\ \hline
\multirow{2}{*}{12} 	& 2 & 2 & 4 & 4 & 0.9167 & 0.9093 & 0.81 \\
			& 4 & 4 & 2 & 2 & 1.4667 & 1.4549 & 0.80 \\ [1ex] 
\hline 
\end{tabular}
\caption{Analysis and Simulation Results [\textit{Mbps}] for multirate AP in 
IEEE 802.11\textit{b}. }
\label{table:aps_stas} 
\end{table}   

In 802.11g, the different possible data rates are 54, 48, 36, 24, 18, 12 and 6 
Mbits/s. Qualnet~4.5 is configured to this mode by setting the channel 
frequency for the 802.11a radio as 2.4 GHz. In Table~\ref{table:multirate_11g}, 
comparisons between analytical and simulation values are given.
\begin{table}[!hH]
\centering 
\begin{tabular}{|c|c|c|c|c|c|c|c|c|c|} 
\hline 
 & \multicolumn{6}{|c|}{No. of STAs at rate (Mbps)} & \multicolumn{3}{|c|}{ Aggregate
 Throughput (Mbps)} \\ \hline
M & 54 & 48 & 36 & 24 & 18 &  6 & Analysis & Simulation & Error\% \\  [0.2ex]
\hline 
\multirow{6}{*}{15} & 1	& 2	& 3	& 4 & 2	    & 3    & 8.14  & 8.18 & 0.49 \\
& 2	& 1     & 3	& 4	& 2     & 3 & 8.16  & 8.20 &  0.48 \\
& 3	& 2	& 1	& 4	& 2	& 3 & 8.31  & 8.32 &  0.12 \\
& 4	& 3	& 2	& 1	& 3	& 2 & 10.22 & 10.25 & 0.29 \\
& 3	& 2	& 4	& 3	& 1	& 2 & 10.38 & 10.41 & 0.29 \\
& 3	& 2	& 4	& 3	& 2	& 1 & 12.27 & 12.31 & 0.24 \\ [1ex]
\hline 
\end{tabular} 
\caption{Throughput [\textit{Mbps}] of multirate AP by analysis and simulation 
for IEEE 802.11\textit{g}. } 
\label{table:multirate_11g} 
\end{table}

\begin{table}[!hH]
\centering 
\begin{tabular}{| c |c | c| c | c | c | c | c|} 
\hline 
 & \multicolumn{4}{|c|}{No. of STAs at rate (Mbps)} & \multicolumn{3}{|c|}{ Aggregate
 Throughput (Mbps)} \\ \hline
M & 11 & 5.5 & 2  & 1 &  Analysis & Simulation & Error \%  \\  [0.5ex]
\hline 
\multirow{2}{*}{10} 	& 2 & 3 & 2 & 3 & 1.1221 & 1.1131 & 0.80 \\ 
			& 1 & 2 & 3 & 4 & 0.8889 & 0.8814 & 0.81 \\ \hline
\multirow{2}{*}{12} 	& 2 & 2 & 4 & 4 & 0.9715 & 0.9637 & 0.81 \\
			& 4 & 4 & 2 & 2 & 1.5647 & 1.5523 & 0.80 \\ [1ex] 
\hline 
\end{tabular}
\caption{Analysis and Simulation Results [\textit{Mbps}] for multirate AP in 
IEEE 802.11\textit{b}. Delayed ACKs are implemented at the receiver with 
alternate segment being acknowledged ($d$ = 2). }
\label{table:aps_stas_d} 
\end{table}

In Tables \ref{table:aps_stas_d}, and \ref{table:multirate_11g_d} comparisons 
between analytical and simulation values are given for TCP with delayed ACKs.
We observe that in all cases, the analytical results are in excellent agreement
with simulations.

\begin{table}[!hH]
\centering 
\begin{tabular}{|c|c|c|c|c|c|c|c|c|c|} 
\hline 
 & \multicolumn{6}{|c|}{No. of STAs with at (Mbps)} & \multicolumn{3}{|c|}{ Aggregate
 Throughput (Mbps)} \\ \hline
M & 54 & 48 & 36 & 24 & 18 &  6 & Analysis & Simulation & Error\% \\  [0.2ex]
\hline 
\multirow{6}{*}{15} & 1	& 2	& 3	& 4 & 2	& 3 & 8.14  & 8.18 & 0.49 \\
& 2	& 1     & 3	& 4	& 2     & 3 & 8.16  & 8.20 &  0.48 \\
& 3	& 2	& 1	& 4	& 2	& 3 & 8.31  & 8.32 &  0.12\\
& 4	& 3	& 2	& 1	& 3	& 2 & 10.22 & 10.25 & 0.29 \\
& 3	& 2	& 4	& 3	& 1	& 2 & 10.38 & 10.41 & 0.29 \\
& 3	& 2	& 4	& 3	& 2	& 1 & 12.27 & 12.31 & 0.24 \\ [1ex]
\hline 
\end{tabular}
\caption{Throughput [\textit{Mbps}] of multirate AP by analysis and simulation 
for IEEE 802.11\textit{g}. Delayed ACKs are
implemented at the receiver with alternate segment being acknowledged ($d$ = 2). } 
\label{table:multirate_11g_d} 
\end{table}

\section{Discussion}\label{sec:Discussion}
In this work, we presented an analytical model to obtain the aggregate throughput
when several TCP-controlled long file downloads are going on. Now let us consider
simultaneous TCP uploads and downloads. The attempt behaviour of nodes
is independent of the packet length. If we interchange downlink data packets sent by
the AP with ACK packets and  uplink ACK packets sent by stations with the TCP 
data packets, the same analysis holds good for the TCP-controlled file uploads.
 
Another case arises when some stations are uploading and some are downloading long
files. Here also our basic Markov model for number of stations with packets to 
send remains the same, if all the TCP windows are equal. Even different window 
sizes can be taken care of by this approach. Some of these extensions have been
analyzed in \cite{astn_model:pradeep_kuri2}.

In our simulation and numerical 
evaluation, we used the 802.11b and 802.11g standards. However, our mathematical 
expressions are independent of these standards, and hence the model can be applied 
to any other standard that has different number of physical data rates.

\section{Conclusion}\label{sec:Conclusion}
In this work, we have presented a simple analytical model for the aggregate 
throughput for TCP-controlled long file transfers in
a ``multirate'' AP. 
We verified the correctness of the analytical model with the simulation results.
As future work, we plan to consider short file transfers.  This can be used to
estimate the delay seen by stations. Further, association schemes can be built 
upon this.


\appendix
\appendixpage

Expressions for probabilities and times discussed in Section~\ref{sec:Analysis}\\
\begin{tabular}{ p{.7cm} p{13cm} }
 $ P_{idle} $        & is the probability of the slot being idle. \\
 	             & $ =  (1 -\beta_{N+1})^{N+1} $ \\
 $ P^{r_i} _{sAP} $  & is probability that the AP wins the channel and transmits a data packet at 
  rate $r_i$ (\textit{i.e.}, to a rate $r_i$ STA). \\
		     &  $ = p_i \beta_{N+1} (1 - \beta_{N+1})^{N}$\\
 $ P^{r_i} _{c} $    & is the probability of the collision event which involves the AP and STAs,
                     with the AP transmitting to a rate $r_i$ STA \\ 
		     & $ = p_i \beta_{N+1} \left(1 - \left( 1 - \beta_{N+1} \right)^{N} \right) $ \\
 $ P^{r_i} _{sSTA} $ &  is the probability that an STA at rate $ r_i $ wins the channel \\
 		     & $ = n_i \beta_{N+1} (1- \beta _{N+1} )^N  $	\\		  
 $ P^{r_i} _{cSTA} $ & is the probability of a collision event involving only
                     STAs at rate $ r_i $ and higher.
 		     Here, the event we are interested in is as follows. One STA at rate $ r_i$ 
                      transmits AND at least one other STA at rate $r_i$ or higher transmits AND no STA 
			at rate $r_{i+1}$  or lower transmits. This  probability is obtained as follows. We consider
			  two cases: (a) 2 or more rate $r_i$ STAs transmit,
                           and we do not care about whether any higher rate  
			   STA transmits or not; (b) Exactly one rate $r_i$ STA  
			  transmits AND at least one higher rate STA
			   transmits. Then, the probability is \\
		& $ \left[ (1-(1-\beta_{N+1})^{n_i} - n_{i} \beta _{N+1} (1- \beta _{N+1})^{n_i -1}) \right. $ \\ 
			  & $ + n_i \beta _{N+1} ( 1- \beta _{N+1} ) ^{n_i -1 } 
			  \left. ( 1- ( 1-  \beta _{N+1}) ^ {n_1+ n_2 +... + n_i -1})  \right] \times $ \\
                & $(1 - \beta_{N+1})^{n_{i+1} + n_{i+2} + \ldots + n_k + 1} $
\end{tabular}\\
To verify that the sum of all the probabilities is 1~:
\begin{equation*}
\fontsize{9}{9} \selectfont
\begin{split}
& P_{idle} + \sum _{i =1 }^{k} P^{r_i} _{sAP} + \sum _{i =1 }^{k} P^{r_i} _{c} + \sum _{i =1 }^{k} P^{r_i} _{sSTA} + \sum _{i =1 }^{k} P^{r_i} _{cSTA} \\
& = (1- \beta _{N+1}) ^{N+1} + \beta _{N+1} (1- \beta _{N+1})^{N} + N \beta _{N+1} (1- \beta _{N+1})^{N}  \\
& +\beta _{N+1} ( 1 - (1- \beta _{N+1})^{N} ) + \sum _{i=1}^{k} (1-\beta _{N+1})^{ n_{i+1} +... +n_{k} +1 } \\
& - \sum _{i=1} ^{k} (1-\beta _{N+1})^{n_{i} +... +n_{k} +1 } -\sum _{i=1}{k} n_i  \beta _{N+1} 
(1-\beta _{N+1})^{N} \\
& =  (1 - \beta _ {N+1} ) ^ {N+1} + \beta_{N+1} (1 + N (1-\beta _{N+1}) ^ N ) \\
& + (1-\beta _{N+1}) - (1-\beta _{N+1})^ {N+1} - N \beta _{N+1}  (1-\beta _{N+1})^ {N+1}\\
& = 1
\end{split}
\end{equation*}
From the above verification, it is clear that all possibilities events have been
considered in Equation~\eqref{eq:enx11}.

\begin{tabular}{ p{.7cm} p{10cm} }
\fontsize{9}{9} \selectfont
$ T^{r_i} _{c} $ &  is the collision duration when the AP and    
			   STAs at rate $ r_i $ are involved. \\
			  & $ =T_p + T_{PHY}+\frac{ L_{MAC}+L_{IPH}+L_{TCP-ACK} }{ r_i } + T_{EIFS}$ \\	  
$ T^{r_i} _{sAP} $  & is the time taken by AP to send a packet to an STA at rate  $ r_i $ \\
			  & $ = T_p + T_{PHY} + \frac{ L_{RTS}}{ C_c } + T_{SIFS} + T_{p}   
			  + T_{PHY} + \frac{ L_{CTS} }{ C_c }  + T_{SIFS} + T_p + T_{PHY} 
			   + \frac{ L_{MAC} + L_{IPH} + L_{TCPH} + L_{TCP} }{ r_i } 
			   + T_{SIFS} + T_{p} + T_{PHY}  + \frac{ L_{ACK} } { C_c } + T_{DIFS} $ \\
$ T^{r_i} _{sSTA} $ &  is the time required to transmit one TCP-ACK 
			  packet from an STA  at rate $ r_i $, including overhead \\
			  & $ = T_p + T_{PHY} + L_{MAC} + \frac{ L_{IPH} + L_{TCP-ACK} }{r_i} 
			   + T_{SIFS} + T_{p} + T_{PHY} + \frac{ L_{ACK} }{ r_i } + T_{DIFS} $ \\ 
$ T^{r_i} _{cSTA} $  &  is the collision duration of STAs at rate $ r_i $ \\
			  & $ = T_p + T_{PHY} + \frac{ L_{MAC} + L_{IPH}+L_{TCP-ACK} } {r_i } + T_{EIFS} $
\end{tabular} \\
The values of $T_p$ , $ T_{PHY} $, $ L_{MAC} $ ,$ L_{IPH} $, $ L_{TCP-ACK} $,
$T_{DIFS}$,$T_{SIFS}$, and $T_{EIFS} $ are standard dependent, and
are mentioned in Kuriakose et al. \cite{astn_model:Kuriakose}, and  Krusheel et al. \cite{astn_model:Krusheel}. 
The values of these parameters are given in Table \ref{table:parameter}.


\begin{table}[hH]
\begin{tabular}{|c|c|c|c|}
\hline
 Parameters     		& Symbol	& 802.11b 	& 802.11g 	\\ \hline \hline
 Max PHY data rate		& $r_d$ 	& 11 Mbps	& 54 Mbps	\\ \hline
 Control rate			& $r_c$ 	& 2 Mbps	& 6 Mbps 	\\ \hline
 PLCP preamble time		& $T_p$ 	& $144\mu s$	& 		\\ \hline
 PHY Header time		& $T_{PHY}$ 	& $48  \mu s$	& $20 \mu s$	\\ \hline
 MAC Header size		& $L_{MAC}$	& 34 bytes	& 34 bytes	\\ \hline
 RTS Header size		& $L_{RTS}$	& 20 bytes	& 20 bytes	\\ \hline
 CTS Header size		& $L_{CTS}$	& 14 bytes	& 14 bytes	\\ \hline
 MAC ACK Header size		& $L_{ACK}$	& 14 bytes	& 14 bytes	\\ \hline
 IP Header size			& $L_{IPH}$	& 20 bytes	& 20 bytes	\\ \hline
 TCP Header size		& $L_{TCPH}$	& 20 bytes	& 20 bytes	\\ \hline
 TCP ACK Packet size 		& $L_{TCP-ACK}$ & 20 bytes	& 20 bytes	\\ \hline
 TCP data payload size 		& $L_{TCP}$	& 1460 bytes	& 1460 bytes	\\ \hline
 System slot time		& $\delta$	& $20 \mu s$	& $9  \mu s$ 	\\ \hline
 DIFS time 			&$T_{DIFS}$	& $50 \mu s$	& $28 \mu s$  	\\ \hline
 SIFS time 			&$T_{SIFS}$	& $10 \mu s$	& $10 \mu s$  	\\ \hline
 EIFS time 			&$T_{EIFS}$	& $364 \mu s$	& $364 \mu s$   \\ \hline
 CWmin				& CWmin 	& 31 		& 15		\\ \hline
 CWmax				& CWmin 	& 1023 		& 1023		\\ \hline
\end{tabular}
\caption{Values of Parameters used in Analysis and Simulation} 
\label{table:parameter}
\end{table} 


\end{document}